# On the effects of size factor on albedo versus wavelength for light scattered by small particles under Mie and Rayleigh regimes

Adrian J. Brown[*1]

[1] SETI Institute, 189 Bernardo Ave, Mountain View, CA 94043, USA

[*] Corresponding author address: Adrian J. Brown, SETI Institute, 189 Bernardo Ave, Mountain View, CA 94043; Tel: +1 (650) 810-0223; email: abrown@seti.org

**POPULAR SUMMARY**

- A model of spectral "bluing" and "reddening" effects caused by particles smaller than the wavelength in the Mie and Rayleigh scattering regimes is presented.
- The conditions under which spectral bluing effects are most effective are explored mathematically.
- In particular, the effect of size of the scatterers is examined so that future observations of this phenomenon can be interpreted with greater confidence.
- Counter-intuitively, spectral bluing and reddening effects occur for all sufficiently monomodal particles at some point in the EM spectrum regardless of optical index $m$, even for large absorption index $\kappa$. The real index $n$ and absorption index $\kappa$ of the particle weakly influence the critical size of the particles to act as bluing or reddening factors.

**ABSTRACT**

Scattering by particles significantly smaller than the wavelength is an important physical process in the rocky bodies in our solar system and beyond. A number of observations of spectral bluing (referred to in those papers as 'Rayleigh scattering') on planetary surfaces have been recently reported, however, the necessary mathematical modeling of this phenomenon has not yet achieved maturity. This paper is a first step to this effect, by examining the effect of grain size and optical index on the albedo of small conservative and absorbing particles as a function of wavelength. The basic conditions necessary for spectral bluing or reddening to be observed in real-world situations are identified. We find that any sufficiently monomodal size distribution of scattering particles will cause spectral bluing in some part of the EM spectrum regardless of its optical index.

# I. INTRODUCTION

Clark et al. [2] recently used data from the VIMS instrument on the Cassini spacecraft currently orbiting Saturn to observe spectral bluing in the visible part of the spectrum. They used the term 'Rayleigh scattering' to describe the observed increase in albedo with decreasing wavelength. In addition, Clark et al. [3] used data from the M-cubed instrument on the Chandrayaan spacecraft to report 'Rayleigh scattering' on the lunar surface. Their abstract suggested a simple model for small particles based on the equations 5.13 and 5.14 of Hapke [4] which is developed further herein. Lucey and Noble [5] attempted to model the spectral reddening (they called this effect 'Space Weathering') albedo of nanophase iron oxide particles on lunar regolith using an approach based on the Hapke bidirectional reflectance model [1, 6]. Brown et al. [7] reported spectral bluing of spectra from the area around the retreating north polar ice cap that could be explained by the models presented below.

The aim of this paper is to examine the mathematical development behind spectral bluing and reddening of monomodal particles with wavelength using Mie and Rayleigh scattering models and examine a range of hypothetical, but physically realizable situations that could be the cause of this phenomenon.

In this study, we explicitly model the single scattering albedo of absorbing particles small compared with the wavelength. This should be distinguished from transmission models of atmospheric Rayleigh scattering of the sky [8-11] and observations of non-absorbing molecular Rayleigh scattering [12]. We ignore the effects of polarization in this paper and intend to return to the wavelength dependence of polarization in future work. Polarization measurements of solar system bodies displaying spectral bluing phenomena are expected to be quite important and significant [9, 13, 14], particularly when measurements of the polarization hemispherical phase function through symmetry maps [15] and through a range of phase angles are available [16, 17].

In this study, in order to remain focused and to draw succinct conclusions, we adopt the following simplifications:

*(a) EM spectrum range limitation*. We consider only the visible to near infrared (VNIR: 0.4-2.5 micron) frequency range – this is not strictly necessary but enables us to address a reduced search space and remain relevant for a large range of astronomical, terrestrial [18], laboratory [19] and planetary science reflectance instruments [20]. The extensions to other parts of the EM spectrum should be rather direct.

*(b) Close packing of regolith*. We consider only Mie and Rayleigh scattering, and we choose to overlook the fact that these are only ideal mathematical descriptions of single scattering particles and do not address the issue of close packing of particles in regoliths, which will inevitably complicate the situation [21, 22]. We justify our approach as an exploratory one – our aim here is to examine the question: "at what point does a particle become scattering, and at what point does if become absorbing, due to its size?". In order to answer this question, we are focusing our attention on the Rayleigh approximation of the scattering process, and ignoring close-packing effects.

*(c) Optical index simplification*. We assume in this study that the optical index $m=n+ i\kappa$ remains constant across the wavelength range we consider here. This is a gross oversimplification, and will introduce difficulties in real-world interpretations, however, in order to isolate the effects of grain size on single scattering albedo, we found it eminently desirable to vary 'only one parameter at a time'. We justify this approach in the following manner: most physical targets in reality will experience an increase in absorption index in the UV/visible due to electronic band transitions. This will make them more absorbing in this critical region, and thus if the grain size of a target determines it to be a reddening agent in the following study, in reality, when variable absorption index is taken into account, this finding will still hold. If a target would be a spectrally bluing agent due to its grain size, having a higher absorption index in the UV/visible region may strongly affect this interpretation. When using the results of this study to estimate the bluing power of a particular material, one might take this into account by using the most pessimistic estimates of $m=n+i\kappa$ in the region of interest (and as shown below, this may not equate to the largest values of $n$ and $\kappa$).

*(d) Monomodal simplification.* We assume in the following study that the target grain size distribution is monomodal. It is a physical fact that this only roughly approximates reality in a limited number of cases. We take this approach because if the size distribution of the targets we are dealing with is broad, then the spectral bluing effect will simply disappear. We intend to develop this argument more quantitatively future work. Bohren and Huffman (p. 107) suggest that the (transmission) bluing effect will be "rapidly vanishing as the dispersion of the particle radii increases" [23]. Therefore, for example, 'blue moons' have been suggested to be caused by a peculiar set of conditions leading to a rare set of atmospheric aerosols [23-25]. Although the mechanism we are studying is different, the comments on size distribution are also appropriate here.

*(e) Particle shape considerations.* Van de Hulst [26] and Kerker [27] in particular have presented models of ellipsoidal particles smaller than the wavelength. We intend to address the effect of particle shape on albedo as a function of wavelength in small particles in future work and restrict this work to homogenous spheres.

All these gross simplifications are hopefully outweighed by the single big idea behind the work in this paper, and that is:

> All particles sufficiently small compared with the wavelength are good absorbers, and that the absorption effects due to size must decrease as the wavelength of light is decreased (holding all other factors constant) and this will cause spectral bluing.

The purpose of this paper is to start to try to understand the limitations on this process and the conditions under which the spectral bluing phenomenon might be observed. As we show below, a sufficiently monomodal size distribution of particles will show spectral bluing at some point in the EM spectrum, regardless of the optical index.

## II. THEORY

### A. Rayleigh scattering approximation

The Clausius-Mossotti relation for relating the electric permeability of a dielectric to its polarizability, α is given by [28-30]

$$N\alpha = \frac{3}{4\pi}\frac{\varepsilon - 1}{\varepsilon + 2} \quad (1)$$

where $N$ is the number density – the number of particles per unit volume, for one particle, $N = \frac{1}{V}$, ε is the dielectric constant and the index of refraction is $m=n+i\kappa$. Using the Maxwell relation $n = \sqrt{\varepsilon\mu} \approx \sqrt{\varepsilon}$ we get this form of the Lorentz-Lorenz relation [31, 32]:

$$\alpha = \frac{3}{4\pi}\left(\frac{n^2-1}{n^2+2}\right)V = \left(\frac{n^2-1}{n^2+2}\right)a^3 \quad (2)$$

where $V$ is the volume and $a$ is the radius of the sphere. We can use the following formulas for the scattering and absorption cross section [26]:

$$C_{sca} = \frac{8\pi}{3}k^4|\alpha|^2 \quad (3)$$

and

$$C_{abs} = 4\pi k \cdot \text{Re}(i\alpha) \quad (4)$$

plugging in (2) into (3) and (4) we get:

$$C_{sca} = \frac{8\pi}{3}k^4\left|\frac{n^2-1}{n^2+2}\right|^2 a^6 = \frac{8}{3}X^4\left|\frac{n^2-1}{n^2+2}\right|^2 a^2\pi \quad (5)$$

$$C_{abs} = 4\pi k \, \text{Im}\left(\frac{n^2-1}{n^2+2}\right)a^3 = -4X \, \text{Im}\left(\frac{n^2-1}{n^2+2}\right)\pi a^2 \qquad (6)$$

where $a$ is the radius of the spheres, $k=2\pi/\lambda$, $X = ka$ is the size factor. We can convert the scattering cross sections into scattering efficiencies, $Q_{sca}$ and $Q_{abs}$, by dividing (4) by $\pi a^2$.

### B. Mie Scattering

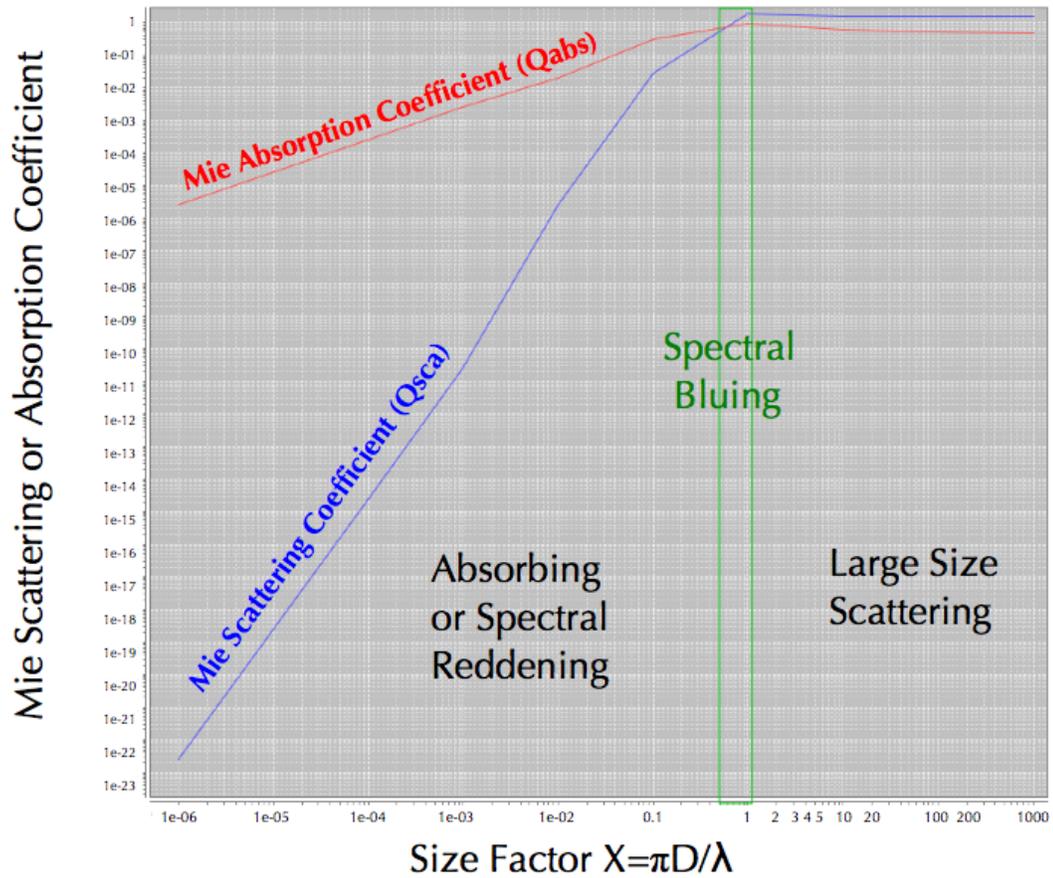

FIG. 1. Adapted from Hapke [1]. We have plotted the Mie scattering and absorption coefficients for a particle with constant optical index $m=3.0+3.0i$. We note the general regions of spectral bluing, reddening and 'Large Size Scattering' based on the results presented in this paper.

Fig. 1 is an adaptation of Fig. 22 from Hapke's treatise on space weathering and the effect of nanophase particles on the albedo of rocky objects [1]. Fig. 1 plots the absorption and scattering coefficients of a Mie target (calculated using the Mie scattering code of Wiscombe [33]) with $m=3.0+3.0i$. We have noted in the left of Fig. 1, where $X < 1$ and the absorption coefficient is decreasing linearly with $X$, the scattering coefficient is decreasing to $X^4$. This results in very low albedo for experiments replicating this physical situation, we call this the region of 'reddening'.

In the central part of Fig. 1, we have marked an area of bluing where the single scattering albedo is increasing with increasing size factor – it is this part of the graph that is the subject of this paper. As we show below, this region is present for all particles, no matter what their optical index, $m$.

**C. Spectral Bluing effects with wavelength for small particles**

Hapke [1] and Lucey and Noble [5] discuss how superfine particles of nanometer size will be absorbing and produced reddened spectra, typically shown in the 'space weathered' surface of the Moon and other airless rocky bodies. However, Clark et al. [2] have noted that it is possible for small particles to scatter more light (increase $Q_{sca}$) in the visible part of the spectrum. How can these two statements be reconciled? There are two ways we mention here.

*(a) Conservative scatterering.* One way in which small particles can scatter efficiently is if they are truly conservative scatterers, i.e. $\kappa=0$. As discussed by Bohren and Huffman on p.132 [23], this is the situation studied by Rayleigh [34] and results in a scattering cross section varying with $\lambda^{-4}$ as in equation (3). Hapke notes on p. 73 [4] that "particles small compared with the wavelength are known as *Rayleigh scatterers*" and on the following page that "small absorbing particles are called '*Rayleigh absorbers*' and Clark [3] has followed this terminology. Bohren and Huffman acknowledge on p.132 that Rayleigh did not address absorbing particles, however they "attach the name 'Rayleigh' to small particle scattering for convenience".

*(b) Absorbing scatterers*. In Fig. 1, we noted a region where it was possible for the scattering coefficient to be higher than the absorption coefficient, where the size factor $X < 1$, this region was marked 'spectral bluing'. We now show that particles where $Q_{abs} < Q_{sca}$ and $X < 1$ have the capacity to produce higher albedo signatures which may appear as bluing of a continuum spectrum.

First we compute the single scattering albedo of a Mie or Rayleigh particle using:

$$w = \frac{Q_{sca}}{Q_{ext}} = \frac{Q_{sca}}{Q_{sca} + Q_{abs}} \qquad (7)$$

as advocated on p. 74 of [4] or p. 123 of [27].

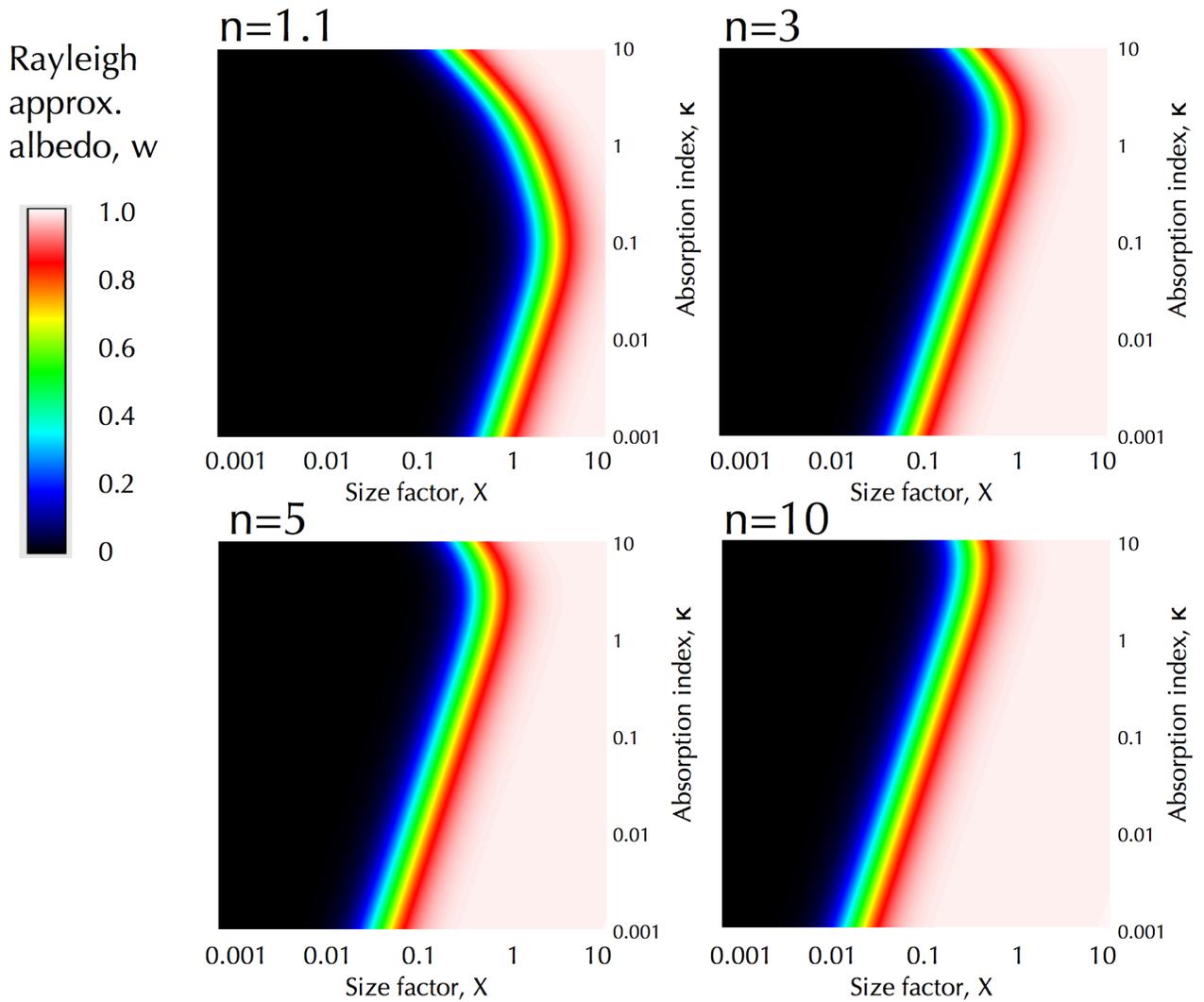

FIG. 2. The Rayleigh approximation of *w*, the single scattering albedo (see equation 6) plotted for varying size factor *X* on the range axis and absorption coefficient, κ, on the domain axis. Note both axes are logarithmic. The left of each plot (small grains) is black, showing regions where small particles are absorbing. Regions of increased value of *w* on the center of the plot track the regions of the search space of *X* and κ where bluing is likely to occur. In the right of the plot, (the 'large grain size' scattering region) is consistently high in the Rayleigh approximation (which doesn't hold in this part of the plot, see discussion on Fig. 4 and 5). Four plots are presented, showing how these results change for variations in the real index of refraction, *n*=1.1, 3, 5 and 10. See text for further discussion.

## III. NUMERICAL RESULTS

### A. Mathematical approach to determining and optimizing Spectral Bluing and Absorbing conditions

*(a) Rayleigh approximation to single scattering albedo.* Substituting equations (5) and (6) into equation (7), we arrive at:

$$w(X,n,\kappa) = \frac{X^3((n-1)^2 + \kappa^2)((n+1)^2 + \kappa^2)}{X^3((n^2+\kappa^2)^2 - 2(n^2-\kappa^2)+1) + 9\kappa n} \quad (8)$$

expanding this expression using a partial fraction decomposition, we get:

$$w(X,n,\kappa) = 1 - \frac{9n\kappa}{X^3((n^2+\kappa^2)^2 - 2(n^2-\kappa^2)+1) + 9\kappa n} \quad (9)$$

Fig. 2 shows the Rayleigh approximation for the single scattering albedo $w$, according to equations (8) and (9). The single scattering albedo is plotted for against size factor $X$ and absorption coefficient, $\kappa$. The left of each plot (smaller grains) is black, showing regions where small particles are absorbing. Regions of increasing value of $w$ on the center of the plot track the regions of the search space of $X$ and $\kappa$ where bluing is likely to occur. In the right of the plot, the Rayleigh approximation is not expected to hold (e.g. for $X > 0.4$, see below), and so the constant values of $w(=1)$ in the plot should be disregarded. Plots for $n=1.1, 3, 5$ and $10$ are presented to show how variations in the real index of refraction affect the Rayleigh approximation of $w$. For increasing $n$, the region of bluing moves further to the left, and the convex bend (at $X=1$, $\kappa=0.1$ for $n=1.1$) moves 'upwards' (to higher $\kappa$) with increasing $n$.

The most important conclusion from Fig. 2 is that, somewhat counter intuitively, reddening and bluing occur for particles of any optical index *m*, even for large absorption index κ. The real index *n* and absorption index κ weakly influence the critical size of the particles where they act as Rayleigh absorbers or Rayleigh scatterers.

We can obtain the third order Taylor's series expansion at $X=0$ for equation (9) as follows:

$$w(X \to 0, n, \kappa) \sim \frac{X^3((n-1)^2 + \kappa^2)((n+1)^2 + \kappa^2)}{9\kappa n} + O(x^4) \quad (10)$$

*(b) Rayleigh approximation to dw/dX.* Differentiating equation (8) once (using the quotient rule) with respect to *X*, we arrive at the following relation:

$$\frac{dw}{dX}(X, n, \kappa) = \frac{(27\kappa n^5 + (54\kappa^3 - 54\kappa)n^3 + (27\kappa^5 + 54\kappa^3 + 27\kappa)n)X^2}{(n^8 + (4\kappa^2 - 4)n^6 + (6\kappa^4 - 4\kappa^2 + 6)n^4 + (4\kappa^6 + 4\kappa^4 - 4\kappa^2 - 4)n^2 + \kappa^8 + 4\kappa^6 + 6\kappa^4 + 4\kappa^2 + 1)X^6 + (18\kappa n^5 + (36\kappa^3 - 36\kappa)n^3 + (18\kappa^5 + 36\kappa^3 + 18\kappa)n)X^3 + 81\kappa^2 n^2} \quad (11)$$

which can be simplified to:

$$\frac{dw}{dX}(X, n, \kappa) = \frac{X^2(27\kappa n((n-1)^2 + \kappa^2)((n+1)^2 + \kappa^2))}{(X^3((n^2 + \kappa^2)^2 - 2(n^2 - \kappa^2) + 1) + 9\kappa n)^2} \quad (12)$$

Fig. 3 shows the first derivative of the Rayleigh approximation (*dw/dX*) according to equation (11) plotted for varying size factor *X* on the range axis and absorption coefficient, κ, on the domain axis. Plots for *n*=1.1, 3, 5 and 10 are presented, showing how these results change for variations in the real index of refraction.

We can interpret the differential graphs as showing us regions of the *n*,κ space that will provide the best opportunities for maximal bluing effects. Those areas where the derivative values are greatest will give the strongest bluing effect.

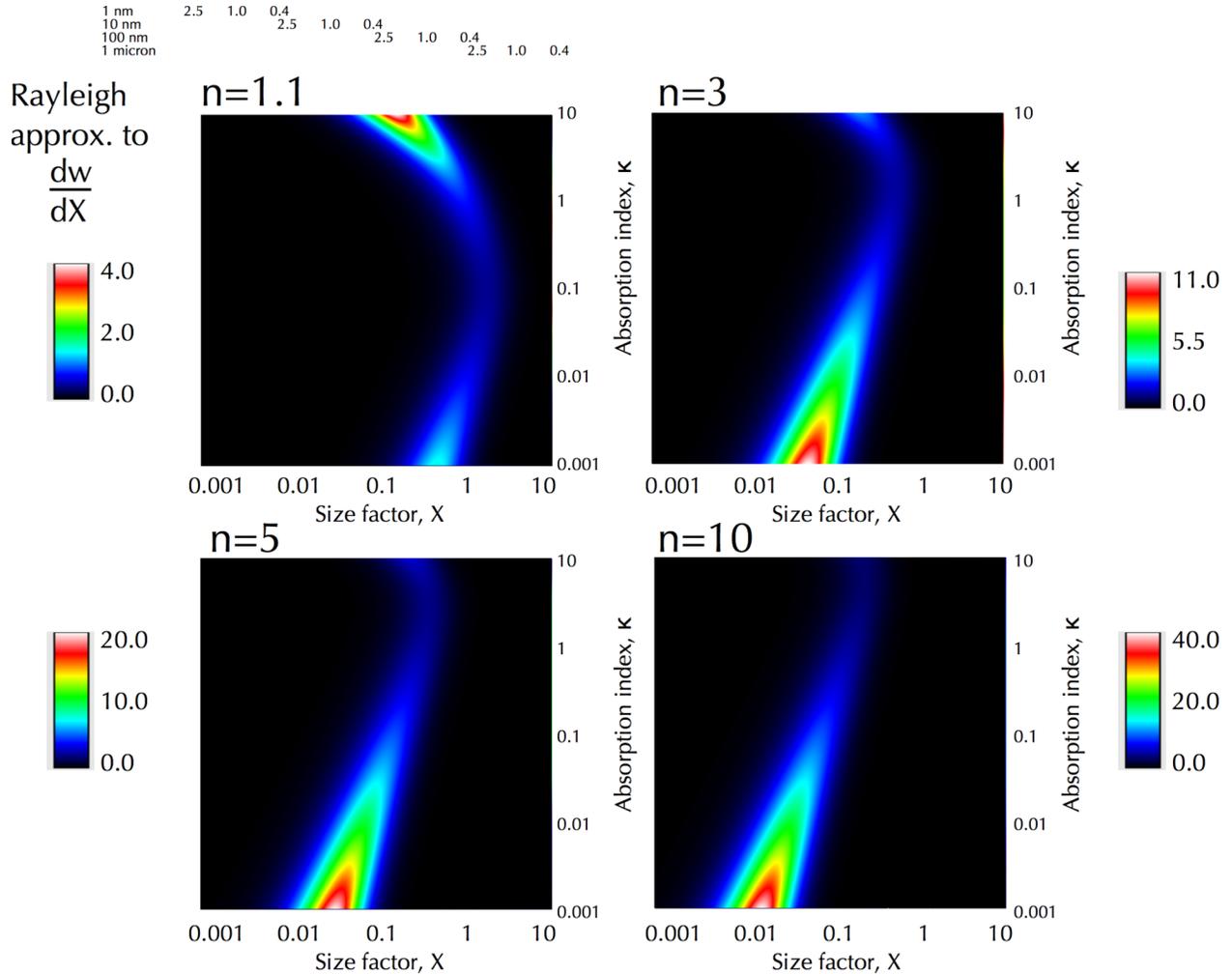

FIG. 3. The first derivative of the Rayleigh approximation to *dw/dX* (equation (11)) plotted for varying size factor *X* on the range axis and absorption coefficient, κ, on the domain axis. Note both axes are logarithmic. The regions of increased value of the first derivative track the regions of the search space of *X* and *κ* where bluing is likely to occur. Four plots are presented, showing how these results change for variations in the real index of refraction, *n*=1.1, 3, 5 and 10. Note the scale of each image is different. A cheat sheet for converting size factor to grain size in VNIR is provided on the lop left. See text for further discussion.

*(c) Small Size Factor approximations.* When $X \ll 1$, an approximation to equation (11) is obtained by setting the $X^6$ term in the denominator to zero:

$$\frac{dw}{dX}(X,n,\kappa) \sim \frac{X^2(27\kappa n^5 + (54\kappa^3 - 54\kappa)n^3 + (27\kappa^5 + 54\kappa^3 + 27\kappa)n)}{(18\kappa n^5 + (36\kappa^3 - 36\kappa)n^3 + (18\kappa^5 + 36\kappa^3 + 18\kappa)n)X^3 + 81\kappa^2 n^2} \quad (13)$$

we can also derive a Taylor's series for *dw/dX* at *X*=0, obtaining

$$\frac{dw}{dX}(X \to 0, n, \kappa) \sim \frac{X^2((n-1)^2 + \kappa^2)((n+1)^2 + \kappa^2)}{3\kappa n} + O(X^5) \quad (14)$$

Equations (13) and (14) approximate the behavior of the *dw/dX* at $X = 0$.

The strength of *dw/dX* grows with increasing *n* from *n*=1.1 to *n*=10 (note the scale of each plot in Fig. 3 is different from the other color scales). *dw/dX* increases in each plot with decreasing $\kappa$, for $\kappa < 0.11$. However, *dw/dX* increases above a certain value of $\kappa$, dependent on the value of *n*. For *n*=1.1, this is around $\kappa$=0.1.

*(d) Mie single scattering albedo calculations.* It should be remembered that Rayleigh approximation is only an approximation to the light scattering process, and thus the above results are not be applicable outside a restricted range of size factors. Penndorf [35] reminds us that the Rayleigh approximation is only expected to hold for *nX* < 1, and Cox et al. [36] report good matches of the Rayleigh approximation to experiment for *nX* ~1.21, but not beyond that point. Kim et al. [37] showed that Rayleigh scattering is a good approximation for Mie scattering for $\lambda < 0.4X$. Kerker et al. [38] and Ku and Felske [39] have also pointed out the limitations of the Rayleigh scattering approximation. For this reason we also calculate the Mie absorption and scattering coefficients in this paper, with the understanding that spherical scatterers are also likely to be an over-simplification of the real world physical situation (e.g. on planetary surfaces).

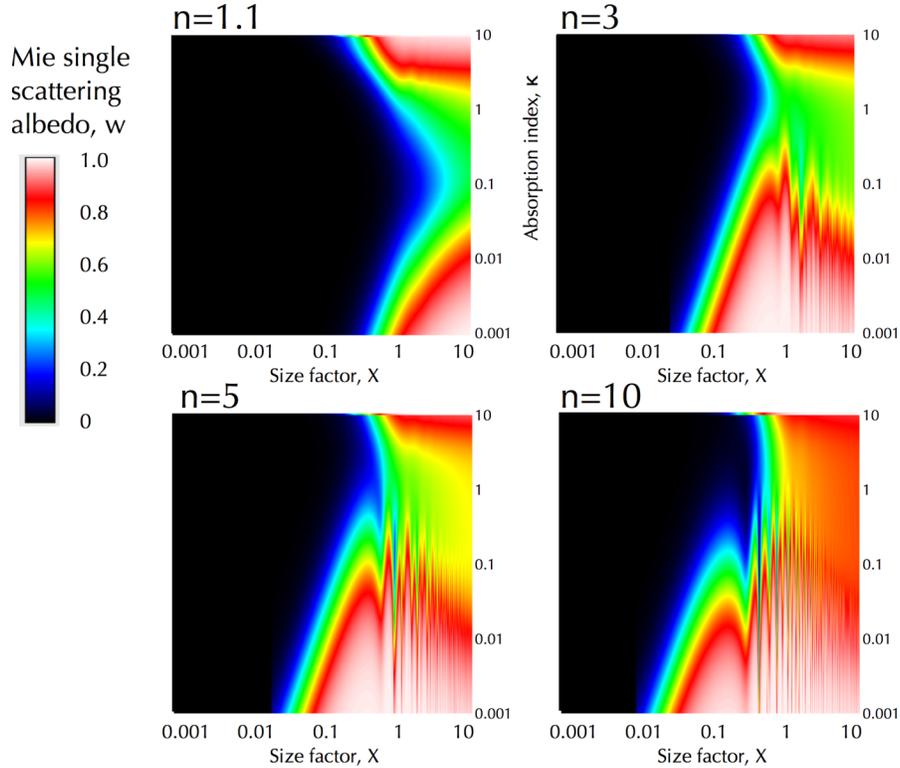

FIG. 4. Mie theory calculations of single scattering albedo for varying index of absorption κ and size factor, *X*. Four instances of real index *n*=1.1, 3, 5 and 10 are presented. See text for further discussion.

Fig. 4 presents the calculated Mie single scattering albedo using equation (7) and the Mie scattering code of Wiscombe [33] for absorption coefficient κ versus size factor *X*. In Fig. 5 we present a comparison with the Rayleigh approximation by calculating the absolute difference between the Rayleigh (Fig 2.) and Mie (Fig. 4) solutions. This primarily highlights the difference between the two solutions due to Mie resonances at *X* ~ 1.

It is of particular interest to this study that there is very little difference in the bluing region (the slope up from 0 from left to right around *X* ~ 1). Small particles that are absorbing due to their size in the Rayleigh approximation and also absorbing in the Mie theory, and although there is a decrease in the magnitude of the albedos around 0.1 < *k* < 1 (particularly apparent in the top two images for *n*=1.1 and *n*=3, and also in the bottom left for *n*=5) the region of increasing albedo (the spectral bluing region) is almost identical. For this reason, we feel confident in interpreting our Raleigh derivative results in Fig. 3 as the regions of most effective spectral bluing, which is the key finding of this paper.

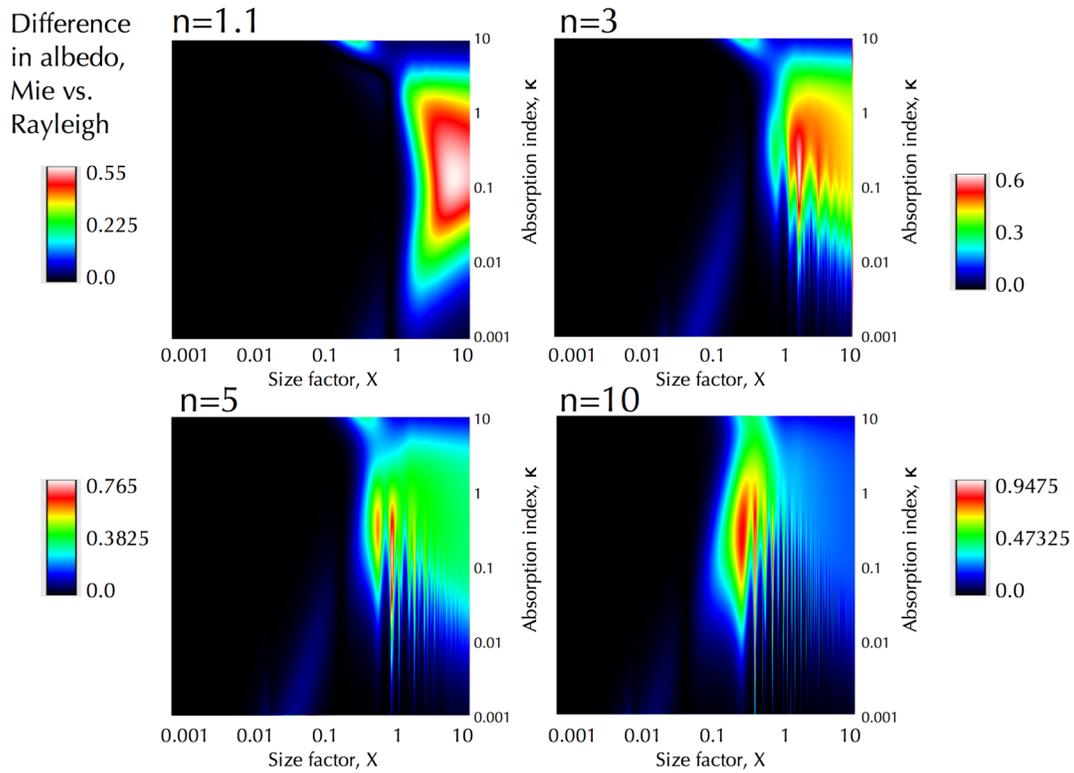

FIG. 5. The calculated absolute difference between the Rayleigh approximation and Mie single scattering albedos for varying index of absorption κ and size factor $X$. Note the scales of each plot are different. Most of the observed difference is caused by the Mie resonances around $X \sim 1$. See text for further discussion.

**B. Predictions and Conditions for achieving bluing in the VNIR**

With the preceding discussion, we are now in a position to make some predictions regarding the appearance of bluing in the VNIR as reported by Clark et al. [2, 3] and Brown et al. [7]. We interpret the observed bluing reported in these three cases by trying to approximately fit the bluing range in each case. We do not attempt to numerically fit the data, but give a qualitative prediction for the grain size range that would best fit the observations given th simplifications we have already discussed.

In order to make a reasonable prediction of the most likely grain size responsible for spectral bluing, we make a further simplifying assumption that the visible optical index of the small particles is in the range $1.1 < n < 3$ and $0.03 < \kappa < 3$, because in this region the bluing is limited to a relatively restricted band of size factors (from $0.5 < X < 1.2$, see Fig. 3). This covers a large portion of naturally occurring materials.

We have created a plot showing the span of the band of $0.5 < X < 1.2$ for a range of grain diameters, $0.1 < D < 1$ (microns) and wavelengths $0.2 < \lambda < 2.5$ (microns), this is shown as Fig. 6. For this restricted range of optical indexes, Fig. 6 provides an estimate at the most range of most effective grain diameters to produce bluing.

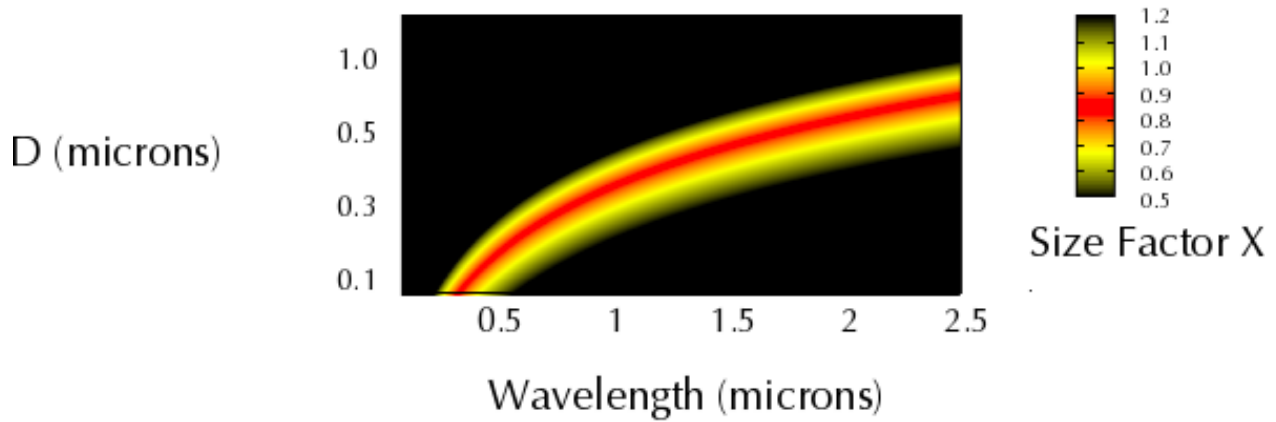

FIG. 6. An estimate of the most effective grain diameter for bluing at particular wavelengths in the VNIR wavelength range. **NOTE: this relationship only holds for optical indexes in a restricted range $1.1 < n < 3$ and $0.03 < \kappa < 3$.** Because the size factor range $0.5 < X < 1.2$ is most effective for bluing (see Fig. 3, particularly the top two panels) in this range of optical indexes, we have highlighted this range of $X$ in red and yellow. To estimate the grain diameter range that is most effective at any wavelength in the VNIR, one may take the limits of this band for any particular $\lambda$ in microns. See text for further discussion.

We will now use the results in Fig. 6 to interpret three Rayleigh scattering measurements reported to date in the literature.

*(a) Lunar regolith.* Clark et al. [3] reported a steep increase in reflectance in the visible range (starting at 1 μm and increasing most steeply at 0.5 μm) using spectra from the Chandrayaan/M^3 data. Based on the range of the observed spectral bluing effect, and making the assumptions discussed above, we can bracket the most likely grain diameter of the scatterers between 0.1 and 0.5 micron, with best fits around 0.3 microns.

*(b) Saturnian icy satellites.* Clark et al. [2] reported a steep increase in reflectance in the visible range at Dione (from 0.4-1.0 μm) and a more restricted blue peak for Epimetheus (from 0.4 to 0.7 μm) for data collected from the Cassini VIMS instrument. Based on these observations, and with the assumptions stated above, the grain diameter of the small scatterers on Dione would best fit 0.3 microns, and must lie within the 0.1-0.5 micron range. Epimetheus grain diameter would best fit 0.2 microns, and must lie in the 0.05-0.4 micron range.

*(c) Martian regolith.* Brown et al. [7] reported a number of VNIR spectra from the CRISM instrument of Martian soil/regolith (observed after the winter polar $CO_2$ ice cap had recently disappeared) that showed a gradual increase in reflectance from 1-2μm. Based on the assumptions listed above, a grain diameter of 0.5 microns would best fit the data, and the scattering particles must lie within the range of 0.3 to 0.8 microns in order to produce the observed effect.

## IV. DISCUSSION

One might ask - what is the best use of Fig. 2 and 3 showing $w$ and $dw/dX$ for $k$ vs. $X$? The answer is that they show regions where bluing will be most effective. It should be noted that weak regions of bluing might be possible in regions where the $dw/dX$ is small, however the most effective spectral bluing regions are shown as maxima in Fig. 3. It should be remembered that these particles are likely observed with other 'non-bluing' agents (within the field of view, or intimately associated with the bluing grains) so the effect of bluing will be diminished, thus making it important that the scattering agents possess strong bluing capabilities.

We wish to mention a couple of related studies of small particle scattering and mention how they differ from this study.

*(a) Blue suns and Moons.* As mentioned above and discussed by Wilson [40], Pesic [11] and Adler and Locke [41], the phenomenon of blue suns and blue moons is explained well as a transmission effect that is related to variations in the Mie scattering cross section of a small particle due to Mie resonances, and the particles sizes we have studied here (with $X < 1$) are too small to achieve this effect. A simple diffraction-based model for blue moons and blue suns has been suggested by Pesic [11], however he considered the effects of transmission through a cloudy atmosphere and the effects of decreasing $Q_{ext}$ over various ranges of size factor $X$. Instead, in this paper we have considered the spectral bluing effect by concentrating on the variation of the single scattering albedo of the small particles.

*(b) Rayleigh scattering near terrestrial clouds*. 'Rayleigh scattering' of reflected sunlight near clouds has also been of some concern in the terrestrial environments. Most suggested explanations for this effect suggest molecular Rayleigh scattering above cloud tops in gaps between clouds might be responsible for this effect [42, 43].

A similar, but angular-dependent, coloration of the spectrum occurs when light is reflected in the specular direction from a randomly rough surface, as discussed in [44].

*(c) Scattering by small absorbing particles*. Many studies have been made of the variable absorption properties of small particles using Mie theory. Deirmendijian [45], Plass [46] and Kattawar and Plass [47] provided the first extensive investigations into scattering by highly absorbing spheres using advanced Mie scattering codes, and emphasized the variations in $Q_{sca}$ and $Q_{ext}$ as a function of size factor $X$, with particular attention on the patterns of Mie resonances. Faxvog and Roessler [48] conducted a similar study of absorbing Mie spheres and found the grain diameter for greatest reduction in visibility for aerosols of carbon ($m$=1.96+0.66i), iron (m=3.51+3.95i), silica (m=1.55) and water ($m$=1.33).

Finally, Mishchenko et al. [49] studied the effect of a 'dusting' of sub-microscopic particles on wavelength size particles using a numerically exact T-matrix formulation. They calculated scattering efficiencies as a function of scattering angle and found that the small particles is less significant than a major asphericity of the scattering object. None of these studies touched on the potential for spectral bluing for particles small with the wavelength.

## V. CONCLUSIONS

We have investigated the theoretical dependence of Rayleigh and Mie single scattering albedo of small particles as a function of grain size for a wide range of the optical index $m$. Our findings can be summarized thus:

- Bluing and reddening effects are predicted by Mie and Rayleigh theory and that the single scattering albedo of any isolated particle with $X < 1$ is strongly determined by its size and weakly by its optical index, $m$.

- All sufficiently monomodal scatterers will cause spectral bluing at some point in the EM spectrum regardless of their optical index $m$.

- Counter intuitively, spectral bluing and reddening effects occur for particles of any optical index $m$, even for large absorption index $\kappa$. The real index $n$ and absorption index $\kappa$ weakly influence the critical size of the particles where they act as bluing or reddening agents.

- The region of spectral bluing (the region where the Rayleigh approximation to $w$ (equation (9)) is increasing rapidly with increasing size factor) varies weakly with optical index $m$.

- We have presented a mathematical relation to describe the conditions of maximal effectiveness of spectral bluing (equation (12)). In order to achieve a bluing effect in the VNIR part of the spectrum, for a large range of optical index found in natural settings ($1.1 < n < 3$ and $0.03 < \kappa < 3$) size factors from $0.5 < X < 1.2$ provide the most effective bluing capability (Fig. 3).

- We have applied the results of this investigation to the cases of Rayleigh scattering known to us at this time – on Dione and Epimethius, the Moon and Mars. Using the data presented in Fig. 6, future observations of Rayleigh scattering can now be interpreted with greater confidence.

## ACKNOWLEDGMENTS

This work was partly supported by a grant from the NASA Planetary Geology and Geophysics (PGG) program run by Dr. Mike Kelley. We thank Roger Clark for bringing this fascinating phenomenon to our attention.